\documentstyle[12pt]{article}

\newcommand{\sect}[1]{\setcounter{equation}{0}\section{#1}}

\newcommand{\EQ}{\begin{equation}}
\newcommand{\EN}{\end{equation}}
\newcommand{\bea}{\begin{eqnarray}}
\newcommand{\ena}{\end{eqnarray}}

\renewcommand{\a}{\alpha}
\renewcommand{\b}{\beta}
\renewcommand{\c}{\gamma}
\renewcommand{\d}{\delta}

\newcommand{\la}{\lambda}

\newcommand{\shalf}{\frac{1}{2}}
\newcommand{\pa}{\partial}
\newcommand{\dz}{\frac{dz}{2\pi i}}
\newcommand{\ra}{\rangle}
\newcommand{\lan}{\langle}

\begin{document}

\topmargin 0pt
\oddsidemargin 5mm

\renewcommand{\Im}{{\rm Im}\,}
\newcommand{\NP}[1]{Nucl.\ Phys.\ {\bf #1}}
\newcommand{\AP}[1]{Ann.\ Phys.\ {\bf #1}}
\newcommand{\PL}[1]{Phys.\ Lett.\ {\bf #1}}
\newcommand{\NC}[1]{Nuovo Cimento {\bf #1}}
\newcommand{\CMP}[1]{Comm.\ Math.\ Phys.\ {\bf #1}}
\newcommand{\PR}[1]{Phys.\ Rev.\ {\bf #1}}
\newcommand{\PRL}[1]{Phys.\ Rev.\ Lett.\ {\bf #1}}
\newcommand{\PTP}[1]{Prog.\ Theor.\ Phys.\ {\bf #1}}
\newcommand{\PTPS}[1]{Prog.\ Theor.\ Phys.\ Suppl.\ {\bf #1}}
\newcommand{\MPL}[1]{Mod.\ Phys.\ Lett.\ {\bf #1}}
\newcommand{\IJMP}[1]{Int.\ Jour.\ Mod.\ Phys.\ {\bf #1}}
\newcommand{\JP}[1]{Jour.\ Phys.\ {\bf #1}}
\renewcommand{\thefootnote}{\fnsymbol{footnote}}

\begin{titlepage}

\setcounter{page}{0}
\rightline{OS-GE 22-91}
\rightline{Brown-HET-844}

\vspace{1.5cm}
\begin{center}
{\Large Spectrum of Two-Dimensional (Super)Gravity\footnote{An expanded
version of the talk given at the workshop on {\em Developments
in String Theory and New Field Theories}, Yukawa Institute for
Theoretical Physics, Kyoto University, September 9--12, 1991.}}
\vspace{1.5cm}

{\large Katsumi Itoh$^{1}$ and Nobuyoshi Ohta$^{2}$} \\
\vspace{1cm}
$^1${\em Department of Physics, Brown University \\
    Providence, RI 02912, U. S. A.}\\
\vspace{.5cm}
$^2${\em Institute of Physics, College of General Education,
Osaka University  \\  Toyonaka, Osaka 560, Japan} \\
\end{center}
\vspace{15mm}
\centerline{{\bf{Abstract}}}

We review the BRST analysis of the system of a (super)conformal matter
coupled to 2D (super)gravity. The spectrum and its operator realization
are reported. In particular, the operators associated with the states of
nonzero ghost number are given. We also discuss the ground ring
structure of the super-Liouville coupled to ${\hat c}=1$ matter.
In appendices, hermiticities, states for $c<1$ conformal matter
coupled to gravity and the proof for the spectrum are discussed.

\begin{center}
Submitted to: Progress of Theoretical Physics Supplement
\end{center}

\end{titlepage}
\newpage
\setcounter{footnote}{0}

\sect{Introduction}

String theories, or more generally, 2D conformal field theories (CFTs)
have been studied for their applications to 2D statistical systems and
unified theories including gravity. In their long history, no one has
ever been able to go beyond the perturbative treatment of strings. It
is crucial, however, to understand nonperturbative aspects of the
theories for their applications. The recent discovery of the double
scaling limit in the matrix models~\cite{BDG} has opened a way to
address this problem and has attracted much attention.

The 2D gravity coupled to matter is a system simple enough to be exactly
solvable while retaining many of important features of physical
interest. The key of its solvability is the fact that the model has very
small number of degrees of freedom. The one-dimensional matrix
model~\cite{PGB,JEV} corresponds to interacting strings in
``two-dimensional space-time", where there is no transverse degree of
freedom. One may then naively expect that they describe only the motion
of ``center of mass". Surprisingly enough, it turns out that the system
carries unexpectedly rich dynamical structure with infinite number of
``extra discrete states" other than the ``center of mass" degree of
freedom~\cite{GKN,POL}.

In the discrete (matrix model) approach, it is not clear what conditions
characterize physical states and the origin of these ``extra states"
is obscure. Thus it is very important to understand these results in
the continuum approach where physical state condition is well-defined.
In the conformal gauge, the 2D gravity coupled to a CFT appears as
Liouville theory~\cite{POL1,DDK}. In this formulation physical states
may be specified as the nontrivial cohomology classes of the BRST
operator and the raison d'\^{e}tre of such ``extra states" has been
understood in the BRST formalism~\cite{LIZ,BMP}. However, their roles
have not been fully clarified yet. Some of their known features are the
following: they are the remnants of the higher massive modes of critical
strings; the discrete states are responsible for the presence of a
large algebra, $W_{\infty}$, in the system~\cite{W,KP}.

In our earlier paper~\cite{IOH}, we have reported our results on the
spectrum of $N=1$ supersymmetric CFTs coupled to 2D supergravity (The
same results are also announced in Ref.~\cite{BMP2}). This study is also
important for the following reason. Our present understanding of bosonic
2D gravity is based on the results of two completely different approaches:
discrete (matrix models, collective field theories) and continuum
approaches. Both of them have brought us almost a consistent picture for
the bosonic theory. However, we do not know even how to formulate
supersymmetric theories in matrix models (some related works have been
reported in \cite{JRO}).

Our results are quite parallel to the bosonic case. In the spectrum, we
find massless states in the Neveu-Schwarz (NS) and Ramond (R) sectors,
and discrete states at the levels where we have ``null" states in the
minimal models. In the appendix, we summarize our proof emphasizing
common features of two sectors. One may clearly understand these
discrete states on the basis of two structures: quartet representations
of the BRST formalism~\cite{KUO}; the singular and cosingular vectors
of the Feigin-Fuchs realization~\cite{FF,KMA}. The appearance of the
discrete states is due to ``decomposition of quartets". This has been
reported in our paper and is not repeated here. In this article,
however, we will give a comprehensive review of
the subject together with some new results.

This paper is organized as follows. In the next section, we
review how Liouville theory emerges when 2D
gravity couples to a conformal matter. In sect.~3, we describe the
spectrum of bosonic theories. The results are due to Refs.
\cite{LIZ,BMP}; a summary of the BRST analysis is given in appendix C.
We introduce some new operators with nonzero ghost numbers to construct
discrete states. The $N=1$ supersymmetric case is reported in sect.~4,
emphasizing the similarity to the bosonic case. Sect.~5 is devoted to
discussions. In appendix A, we discuss hermiticity properties of the
Fock space, which should be important in order to discuss the unitarity
but is rarely paid attention. In appendix B, we show how to construct
physical states for $c^M<1$ CFTs coupled to gravity. Finally in
appendix C, we give a description of our analysis of the spectrum which
is relevant to sects.~3 and 4.

\sect{2D gravity coupled to CFT}

In this section, we briefly describe the system of 2D gravity coupled
to a CFT and argue that the gravity sector could be treated as Liouville
theory with the free field measure~\cite{DDK,DHO}.

The partition function of this system is given by
\EQ
Z=\int \frac{{\cal D}g{\cal D}_g X}{V(Diff)}e^{-S(X,g)},
\EN
where $S(X,g)$ is an action for the matter (generically denoted as $X$)
coupled to 2D metric $g$. $S(X,g)$ is invariant under the diffeomorphism
as well as Weyl rescaling of the metric $g\to e^{\sigma}g$
\EQ
S(X,e^{\sigma}g)=S(X,g).
\EN
However, the measure is not Weyl-invariant~\cite{POL1,FRI} and we find
\EQ
{\cal D}_{e^\sigma g}X=e^{(c^M/48\pi)S_L(\sigma,g)}{\cal D}_gX,
\EN
where $c^M$ is the central charge of the matter system
and $S_L$ is the Liouville action
\EQ
S_L(\sigma,g)=\int d^2\xi\sqrt{g}(\shalf g^{ab}\pa_a\sigma\pa_b\sigma+
R\sigma+\mu e^\sigma),
\EN
with $R$ and $\mu$ being the scalar curvature and cosmologial constant,
respectively.

Upon fixing the diffeomorphism invariance in the conformal gauge,
we find the path integral measure in (2.1) becomes
\EQ
{\cal D}_g\phi_0{\cal D}_gb{\cal D}_gc{\cal D}_gX,
\EN
where $b$ and $c$ are Faddeev-Popov ghosts and
$\phi_0$ is a variable parametrizing the Weyl rescaling for a metric
\EQ
g=e^{\phi_0}{\hat g}.
\EN
where ${\hat g}$ is a reference metric parametrized by the moduli space.
The integral over the moduli is suppressed. The integration measure over
$\phi_0$ is defined through the induced norm by the metric (2.6)
\EQ
||\d\phi_0||_g^2=\int d^2\xi\sqrt{g}(\d\phi_0)^2
=\int d^2\xi\sqrt{\hat g}e^{\phi_0}(\d\phi_0)^2,
\EN
which depends on $\phi_0$ itself in a complicated manner. The ingenious
ansatz made by \cite{DDK} and later proved in Ref.~\cite{DHO} is
that we can shift the measure and make it independent of $\phi_0$. In
doing this, one gets a Jacobian $J(\phi,{\hat g})\equiv e^{-S}$:
\EQ
{\cal D}_g\phi_0{\cal D}_gb{\cal D}_gc{\cal D}_gX
={\cal D}_{\hat g}\phi{\cal D}_{\hat g}b{\cal D}_{\hat g}c{\cal D}
_{\hat g}X e^{-S(\phi,{\hat g})},
\EN
where ${\cal D}_{\hat g}\phi$ is the free field measure defined by the
norm
\EQ
\int d^2\xi\sqrt{\hat g}(\d\phi)^2.
\EN
The key assumption is that the most general renormalizable form of $S$
compatible with locality and diffeomorphism invariance is simply
\bea
S(\phi,{\hat g})&=&\frac{1}{8\pi}\int d^2\xi\sqrt{{\hat g}}({\hat g}^{ab}
 \pa_a\phi\pa_b\phi-2Q{\hat R}\phi +4\mu' e^{\a\phi}),\nonumber\\
&=&\frac{1}{2\pi}\int d^2z(\pa\phi{\bar\pa}\phi-\frac{1}{2}Q
 \sqrt{\hat g}{\hat R}\phi +\mu' \sqrt{\hat g}e^{\a\phi}),
\ena
which is similar to the Liouville action (2.4).

The unknown coefficients $Q$ and $\a$ are determined if one notices that
the original theory depends only on $g=e^{\a \phi}{\hat g}$ so that it
is invariant under
\EQ
{\hat g}\to e^{\sigma}{\hat g}, \hspace{5mm} \phi\to \phi-\sigma/\a,
\EN
which means
\EQ
{\cal D}_{e^\sigma{\hat g}}(\phi-\sigma/\a){\cal D}_{e^\sigma{\hat g}}b
{\cal D}_{e^\sigma{\hat g}}c{\cal D}_{e^\sigma{\hat g}}X e^{-S(\phi
-\sigma /\a,e^\sigma{\hat g})}
={\cal D}_{\hat g}\phi{\cal D}_{\hat g}b{\cal D}_{\hat g}c
{\cal D}_{\hat g}Xe^{-S(\phi,{\hat g})}.
\EN
However, $(\phi-\sigma/\a)$ on the left hand side is just an integration
variable and we could have called it $\phi$ itself. Viewed this way,
eq.~(2.12) shows that the conformal anomaly for the total system should
vanish~\cite{DDK}! This means that the central charges from three
sectors should add up to zero
\EQ
c_{total}\equiv c^M+c^L-26=0
\EN
where $c^L=1+12Q^2$. Note that the matter CFT no longer has conformal
invariance by itself because of the conformal anomaly~\cite{POL1,FRI},
but that the inclusion of the Liouville theory recovers the invariance.
The other parameter $\a$ is determined by demanding that $g=
e^{\a\phi}{\hat g}$ be invariant, or $e^{\a\phi}$ be a conformal tensor
of dimension $(1,1)$. Since the dimension of this operator is given by
$-\shalf \a(\a+2Q)$, we obtain
\EQ
\a=-Q+\sqrt{Q^2-2}=\frac{\sqrt{1-c^M}-\sqrt{25-c^M}}{2\sqrt{3}}
\EN
in agreement with Ref.~\cite{KPZ}.\footnote{Note that the relation
(2.14) can be written as $Q=-(\frac{\a}{2}+\frac{1}{\a})$, a well-known
relation in the quantum Liouville theory~\cite{LIO}.}
Here an appropriate branch is chosen so as to agree with
the semiclassical limit ($c^M\to -\infty$). Eq.~(2.14) shows that there
is a bound $c^M\leq 1$ in order for this approach to make sense.

The argument can be easily extended to supersymmetric case~\cite{DHK}.
In the BRST formalism to be used in the present paper, the condition
$c_{total}=0$ is equivalent to the nilpotency of the BRST charge.

\sect{Bosonic non-critical strings}

In this section, we discuss the spectrum of bosonic theory in the
BRST formalism.

Let us first describe the free field realization which we will use for
both the matter and gravity sectors. A scalar field $\phi$ has the
following expansion
\EQ
\phi(z) = q-i(p-\la)\ln z+i\sum_{n\neq 0}\frac{\a_{n}}{n}z^{-n},
\EN
with the commutation relations
\EQ
[\a_n, \a_m] = n \d_{n+m,0},\;\;~~[q,p]=i~~.
\EN
The energy-momentum tensor of the system is
\EQ
T = -\frac{1}{2} (\partial \phi )^2 - i \la \partial^2\phi,
\EN
which satisfies Virasoro algebra with the central charge $c= 1-12\la^2$.
(The Liouville theory described in sect. 2 corresponds to choosing $Q=
i\la$.)

We use two scalar fields for the matter and gravity sectors, which will
be distinguished by the superscripts $M$ and $L$.

By introducing the ghost fields $(b,c)$, we obtain the BRST
charge~\cite{KOG,IKU}
\EQ
Q_B = \oint \dz c(z) (T(z) + \shalf T^{bc}(z)),
\EN
where the energy momentum tensor $T$ contains the contributions from the
matter and gravity sectors. The ghosts have the conformal dimensions,
dim.($b$, $c$) $=(2,-1)$. The nilpotency of the charge gives us a
constraint that the total central charge is equal to zero, or
equivalently
\EQ
(\la^M)^2+(\la^L)^2 =-2.
\EN
Note that $\la^M$ is real whereas $\la^L$ is pure imaginary. Accordingly
the momenta $p^M$ and $p^L$ are real and pure imaginary, respectively.

Our problem is to find the quotient space $Ker Q_B /Im Q_B$ in the
direct product space out of Fock modules for the matter, gravity and the
ghost sectors; ${\cal F} (p^M, p^L) \equiv {\cal F}(p^M) \otimes {\cal F}
(p^L)\otimes {\cal F}^{gh}$, where ${\cal F}(p)$ is a Fock module
with momentum $p$. The ghost number is defined so that the physical
vacuum $c_1|0\ra_{gh}$ has $N_{FP}=0$, where $c_n (n \ge 1)$ and $b_n
(n \ge 0)$ vanish on $c_1|0\ra_{gh}$. The {\em relative cohomology} is
defined as the cohomology of $Q_B$ on ${\cal F}(p^M,p^L)\cap Ker(b_0)$.
We denote them as $H^{(*)}_{rel} (*,Q_B)$ where the superscript
indicates the ghost number. The {\em absolute cohomology} is defined on
${\cal F}(p^M, p^L)$.

We have extra physical states, i.e., discrete states when the momenta
for both the matter and gravity sectors take the special values
parametrized by two integers $j$ and $k$ $(jk > 0)$ as
\EQ
p = t_{(j,k)} + \la = \shalf ( j t_+ + k t_-),
\EN
where $t_{\pm}=-\la \pm \sqrt{\la^2 +2}$. Note that $\la^M$ (or $c^M$)
is a free parameter of our system except the constraint $c^M\leq 1$.
$\la^L$ is determined by eq.~(3.5); if we choose $\la^L = i
\sqrt{(\la^M)^2 +2}$, then $t_{\pm}^L = \mp i t^M_{\pm}$.

We find the following results in Refs.~\cite{LIZ,BMP}.

\newpage
{\large\bf Theorem 1}

For given $\la^M$ (or $c^M$), we find the following nontrivial
cohomology classes.

(1) If $j=0$ or $k=0$, $H^{(0)}_{rel}({\cal F}(p^M, p^L), Q_B)
 = {\bf C}$,

(2) If $j,k\in{\bf Z}_+$, $H^{(0,1)}_{rel}({\cal F}(p^M,p^L),Q_B)
 ={\bf C}$,

(3) If $j,k \in{\bf Z}_-$, $H^{(0, -1)}_{rel}({\cal F}(p^M, p^L),
 Q_B) = {\bf C}$. $\bullet$
\vspace{5mm}

We have given a proof of this theorem in appendix C together with the
super case. Some explanations of each case are in order.

The first case (1) corresponds to the ``tachyon" field in
two-dimensional space-time; the on-shell condition is simply $-p^+p^-
=\shalf[(ip^L)^2-(p^M)^2]=0$, which implies the particle is ``massless"
in terms of these shifted momenta $p^{M,L}=t^{M,L}+ \la^{M,L}$ ($p^\pm$
are defined by $p^\pm=\frac{1}{\sqrt 2}(p^M \pm ip^L)$). The states are
at the level zero. For this case, the physical state condition does not
require $j(\neq 0)$ or $k(\neq 0)$ to be an integer, though the momenta
take the form in (3.6) owing to the on-shell condition.

For cases (2) and (3), the states are called discrete states and at the
level $jk$. They carry fixed values of momenta (3.6) for given integers
$j$ and $k$ (with $-p^+p^-=\shalf[(ip^L)^2-(p^M)^2]=jk$),\footnote{Recall
that the Liouville momenta are pure imaginary. Thus the Liouville field
may be considered like a time variable while the matter corresponds to
space.} but there are two states at the same levels with ghost
number $N_{FP}=0,1$ for case (2) and $N_{FP}=0,-1$ for (3). The states
for (2) are associated with the singular vectors in ${\cal F}(p)$ while
those for (2) are related to the cosingular vectors in ${\cal F}(p)$.
This observation also explains the ghost numbers they carry~\cite{IOH}.

When $c^M=1$, the discrete states form $SU(2)$ multiplets, and hence the
structure may be most easily studied. In the matter sector, we have
$SU(2)$ current algebra with level $\kappa=1$ generated by the following
currents
\bea
J^{\pm}(z) &=& : e^{\pm {\sqrt 2}i \phi^M(z)}:, \nonumber\\
J^0(z) &=& \frac{1}{{\sqrt 2}} i \partial \phi^M(z),
\ena
with the operator product expansions (OPEs)
\bea
J^+(z)J^-(w) &\sim& \frac{\kappa}{(z-w)^2}+\frac{2}{z-w}J^0(w), \nonumber\\
J^0(z)J^{\pm}(w) &\sim& \frac{\pm 1}{z-w} J^{\pm}(w), \nonumber\\
J^0(z)J^0(w) &\sim& \frac{\kappa/2}{(z-w)^2}.
\ena
In particular, their integrals satisfy $SU(2)$ algebra and the discrete
states (including some charged vacuum states) form $SU(2)$ multiplets.

It has been shown~\cite{W,KP} that these states exhibit ground ring
structure with the symmetry group of the area-preserving diffeomorphism
\cite{WIN}.

For studying the ground ring as well as some other physical implications
of the discrete states, it would be useful to have explicit
representations which may be summarized as follows:

\vspace{5mm}
{\large\bf Theorem 2}

(1) For $j,k \in {\bf Z}_+$ and $N_{FP}=0$,
\EQ
\Psi^{(-)}_{Jm} (z) = (J^-_0)^{J-m} e^{i {\sqrt 2} J \phi^M(z)}
e^{{\sqrt 2}(1+J) \phi^L(z)},
\EN
where
\EQ
J^-_0\equiv \oint _{C_z} \frac{d \zeta}{2 \pi i} J^-(\zeta).
\EN

(2) For $j,k \in {\bf Z}_+$ and $N_{FP}=1$,
\EQ
{\tilde \Psi}^{(-)}_{Jm} (z)
= (J^-_0)^{J-m-1} \oint _{C_z} \frac{d \zeta}{2 \pi i}
\frac{K(\zeta)}{\zeta -z}e^{i{\sqrt 2}J \phi^M(z)}
e^{{\sqrt 2}(1+J) \phi^L(z)},
\EN
where
\EQ
K(z) \equiv c(z) J^-(z).
\EN

(3) For $j,k \in {\bf Z}_-$ and $N_{FP}=0$,
\EQ
\Psi^{(+)}_{Jm} (z) = (J^-_0)^{J-m} e^{i{\sqrt 2}J \phi^M(z)}
e^{{\sqrt 2}(1-J) \phi^L(z)}.
\EN

(4) For $j,k \in {\bf Z}_-$ and $N_{FP}=-1$,
\EQ
{\tilde \Psi}^{(+)}_{Jm} (z)
= (J^-_0)^{J-m-1} \oint _{C_z} \frac{d \zeta}{2 \pi i}L(\zeta)
e^{i{\sqrt 2}(J-1/2) \phi^M(z)}e^{{\sqrt 2}(3/2-J) \phi^L(z)},
\EN
where
\EQ
L(z) \equiv b(z)e^{-i \phi^M(z)/\sqrt{2}}
e^{- \phi^L(z)/\sqrt{2}}.
\EN
The $SU(2)$ quantum numbers are related to $(j,k)$ as $J\equiv
|\frac{j+k}{2}|,~m \equiv \frac{j-k}{2}$ (these relations will be
explained later). $\bullet$
\vspace{5mm}

The states representing the nontrivial cohomology classes are obtained
by acting the above operators (with $z=0$) on the physical vacuum
$|\la\ra\equiv | \la^M=0\ra\otimes|\la^L=\sqrt{2}i\ra \otimes c_1|0
\ra_{bc}$.  We have used the same notation as in Ref.~\cite{W} for
eqs.~(3.9) and (3.13). The corresponding states were given in terms of
Schur Polynomials in~\cite{BMP2} (see also~\cite{WY}). Note also that
we have introduced a BRST-invariant operator $K(z)$, which carries
$N_{FP}=1$ and the same charge as $J^-(z)$. It is easy to show that the
above operators create the states with $N_{FP}=\pm 1$ in Ref.~\cite{BMP2}
written in terms of the Schur polynomials and that these states are BRST
invariant.\footnote{Using $K(z)$, we find states in (2) associated with
the absolute cohomology, which take the form as $\psi_{J+m,J-m}^{(1)}
+c_0 \psi_{J+m,J-m}^{(0)}$ in the notation of Theorem 2.2 in Ref.
\cite{BMP2}.} The operator $L(z)$ with $N_{FP}=-1$ does create
appropriate states, but its BRST transformation property is less obvious.

The relations of $J$ and $m$ to $(j,k)$ may be obtained as follows.
First note a state created by (3.9) carries momenta $(p^M, p^L-i\sqrt{2})
=({\sqrt 2}m,-i{\sqrt 2}(1+J))$. On the other hand, we have $t^M_\pm
=\pm{\sqrt 2}$ and $t^L_\pm=-i{\sqrt 2}$. Combined with (3.6),
we find $J$ and $m$ given in the theorem.

The above expressions clearly tell us that operators with $J$ form
$(2J+1)$-plet if $N_{FP}=0$, while operators with the same $J$ but with
$N_{FP}=\pm 1$ form $(2J-1)$-plet with respect to $SU(2)$.

Unfortunately
the operators with $N_{FP}= \pm 1$ are not quite symmetric, but the
corresponding states are dual to each other in the sense that they
have the non-zero inner products defined in appendix A.

In appendix B we show how to relate the states given in
theorem 2 to those for $c^M<1$ CFTs coupled to gravity.
This is a way to find representatives for the cohomology
classes in theorem 1 for a general $c^M<1$.

\sect{$N=1$ supersymmetric non-critical strings}

The structure of the spectrum for $N=1$ supersymmetric case is quite
similar to the bosonic case, although the proof is slightly more involved.
In appendix C, we give a short summary of the proof treating NS and R
sectors at the same time.

Let us introduce some notation for a scalar supermultiplet $(\phi,\psi)$
to realize superconformal algebra. The scalar field $\phi$ has the
same properties as the previous section.\footnote{We will use the same
notation as the previous section if quantities are naturally extended
to supersymmetric case.} The field $\psi$ has the following expansion
\EQ
\psi (z) = \sum_n \psi_{n+\d} z^{-n-\d-\frac{1}{2}},
\EN
with the commutation relations
\EQ
\{\psi _{n+\d},\psi_{m-\d}\} = \d_{n+m, 0},
\EN
where $\d=1/2~(0)$ for NS (R) sector. The super-stress tensors of the
system are
\bea
T & = & -\frac{1}{2} (\partial \phi )^2 -\frac{1}{2} \psi \partial
\psi - i \la \partial^2\phi , \nonumber\\
S & = & i \psi \partial \phi - 2\la \partial \psi,
\ena
which form the $N=1$ superconformal algebra with the central charge
$c= \frac{3}{2} {\hat c} = \frac{3}{2} (1-8 \la^2)$.

We use two sets of the supermultiplets for the matter and gravity
sectors, which will be distinguished by the superscripts $M$ and $L$.

By introducing the ghost fields $(b,c)$ and $(\b, \c)$, we obtain
the BRST charge~\cite{OHT,MIT,ITO}
\EQ
Q_B = \oint \dz [c(z)(T(z)+\shalf T^{gh}(z) )
- \shalf \c (z)(S(z) +\shalf S^{gh}(z)) ],
\EN
where the super-stress tensors $T$ and $S$ contain the contributions
from the matter and gravity sectors. The ghosts $(b,c)$ and $(\b,\c)$
have the conformal dimensions $(2,-1)$ and $(\frac{3}{2},-\shalf)$,
respectively.  In particular, $\c_{n+\d} ^{\dagger} = \c_{-(n+\d)}$,
$\b_{n+\d} ^{\dagger} =- \b_{-(n+\d)}$ and $[\c_{n+\d}, \b_{m-\d}] =
\d_{n+m,0}.$ The nilpotency of the charge gives us a constraint that the
total central charge is equal to zero, or
\EQ
(\la^M)^2+(\la^L)^2 =-1.
\EN

The ghost number is defined so that the physical vacuum $c_1|0\ra_{gh}$
has $N_{FP}=0$, where $c_n (n \ge 1)$, $b_n (n \ge 0)$, $\c_n (n > 0)$
and $\b_n (n \ge 0)$ vanish on $c_1|0\ra_{gh}$. The {\em relative
cohomology} $H^{(*)}_{rel} (*,Q_B)$ is defined as the cohomology of
$Q_B$ on ${\cal F}(p^M, p^L) \cap Ker(b_0) [\cap Ker (\b_0)]$ for NS [R]
sector. The {\em absolute cohomology} is defined on ${\cal F}(p^M, p^L)$.

We have obtained the following results in Ref.~\cite{IOH}
on the spectrum for super-Liouville theory coupled to ${\hat c}\leq 1$
superconformal matter.

\vspace{5mm}
{\large\bf Theorem 3}

For a given $\la^M$ (or $c^M$), we find the following nontrivial
cohomology classes:

(1) If $j=0$ or $k=0$, $H^{(0)}_{rel}({\cal F}(p^M, p^L), Q_B) = {\bf C}$,

(2) If $j,k \in {\bf Z}_+$, $H^{(0, 1)}_{rel}({\cal F}(p^M,p^L),Q_B)
 ={\bf C}$,

(3) If $j,k \in {\bf Z}_-$, $H^{(0, -1)}_{rel}({\cal F}(p^M, p^L), Q_B)
= {\bf C}$,\\   where $j-k \in 2{\bf Z}$ for NS and $\in 2{\bf Z}+1$
for R sectors respectively. $\bullet$
\vspace{5mm}

The momenta, appearing in the theorem, take the same form as (3.6) with
slight modifications in definitions of $t_{\pm}; t_\pm =-\la\pm\sqrt{
\la^2 +1}$. Again if we choose $\la^L = i \sqrt{(\la^M)^2 +1}$, then
$t_{\pm}^L =\mp i t^M_{\pm}$. The fermionic vacuum in the R sector is
determined by the physical state condition. This theorem is proved in
appendix C.

The first case in the NS sector again corresponds to the ``tachyon"
(actually massless) field in two-dimensional space-time.
The states in the R sector are expeted to give space-time fermion.
The states are at level zero for (1). The on-shell
condition for both sectors is simply $-p^+p^-=\shalf[(ip^L)^2-(p^M)^2]
= 0$, which implies the particles are ``massless" in terms of the shifted
momenta $p=t+\la$. Given this degeneracy in the NS and R sectors, it
would be interesting to address the question of space-time supersymmetry.

For cases (2) and (3), the states are called discrete states and at the
level $\shalf jk$. From the on-shell condition, we have $-p^+p^-=\shalf[
(ip^L)^2-(p^M)^2]=\shalf jk$.
The states for (2) is associated with singular vectors in ${\cal F}
(p)$ while those for (3) is related to cosingular vectors in
${\cal F}(p)$. At this level of conceptual understanding,
the reason why we have these extra states is the same
for the bosonic and $N=1$ supersymmetric theories.

When ${\hat c^M}=1$, the discrete states appear in $SU(2)$ multiplets.
In the matter sector, we again have $SU(2)$ current algebra with level
$\kappa=2$ generated by the following currents:
\bea
J^{\pm}(z) &=& :{\sqrt 2}\psi^M(z) e^{\pm i \phi^M(z)}:, \nonumber\\
J^0(z) &=& i \partial \phi^M(z),
\ena
with OPEs in (3.8) with $\kappa=2$. Their integrals satisfy
the $SU(2)$ algebra and the discrete states (including some charged
vacuum states) form $SU(2)$ multiplets.  A difference from the bosonic
case is that the states belong to different sectors according to their
spins; states in the NS sector have integer spins while those in the R
sector have half-integer spins. Like the bosonic $c^M=1$ case, we expect
that there is a ground ring, which is expected to be generated by the
states with $(j,k)=(-1,-2)$ and $(-2,-1)$ in the R sector.

For the NS sector, we find the following representatives in the $q=-1$
picture:

\vspace{5mm}
{\large\bf Theorem 4}

(1) For $j,k \in {\bf Z}_+$ and $N_{FP}=0$,
\EQ
\Psi^{(-)}_{Jm} (z) = (J^-_0)^{J-m}e^{iJ \phi^M(z)} e^{(1+J) \phi^L(z)}.
\EN

(2) For $j,k \in {\bf Z}_+$ and $N_{FP}=1$,
\EQ
{\tilde \Psi}^{(-)}_{Jm} (z)
= (J^-_0)^{J-m-1} \oint _{C_z} \frac{d \zeta}{2 \pi i}
\frac{K(\zeta)}{\zeta -z}e^{iJ \phi^M(z)} e^{(1+J) \phi^L(z)},
\EN
where
\EQ
K(z) \equiv [\shalf \c(z) + c(z) \psi^M(z)]e^{-i\phi^M(z)}.
\EN

(3) For $j,k \in {\bf Z}_-$ and $N_{FP}=0$,
\EQ
\Psi^{(+)}_{Jm} (z) = (J^-_0)^{J-m}e^{iJ \phi^M(z)} e^{(1-J) \phi^L(z)}.
\EN
Here $J\equiv |\frac{j+k}{2}|$ and $m \equiv \frac{j-k}{2}$ are integers.
$\bullet$
\vspace{5mm}

At present, we do not have compact expressions for $j,k \in {\bf Z}_-$
and $N_{FP}=-1$ though their presence is guaranteed by theorem 3. The
states representing the nontrivial cohomology classes are obtained by
acting the above operators on the physical vacuum $|\la\ra \equiv
|\la^M=0\ra \otimes |\la^L=i\ra \otimes c_1 |0 \ra_{bc}\otimes |0
\ra_{\b \c}$, with $\c_{r}|0\ra_{\b\c}=0$ for $r \geq \shalf$.
Note that we have introduced a BRST-invariant operator $K(z)$,
which carries $N_{FP}=1$ and the same charge as $J^-(z)$. These
are the super extension of the operators reported in (3.9)--(3.13).
Again from the above expressions, we see clearly that operators with
$J$ form $(2J+1)$-plet if $N_{FP}=0$, while operators with the same $J$
but with $N_{FP}=1$ form $(2J-1)$-plet with respect to $SU(2)$.

As for the R sector, one must be careful about the fermion zero-modes,
which form two dimensional Clifford algebra. We use representations
$\psi_0^{\pm}\equiv \frac{1}{\sqrt 2}(\psi_0^{M}\pm i \psi_0^{L})
 \equiv \frac{1}{2}(\sigma_1 \pm i \sigma_2)$: the vacua are
two-dimensional spinors. From the condition $F \equiv 2[\b_0, Q_B]=0$,
 we find the spinor structure of the
state $\left( \begin{array}{c} 0 \\ 1 \end{array} \right)$ for cases
(1) and (2) and $\left( \begin{array}{c} 1 \\ 0 \end{array} \right)$
for case (3) in the theorem [see (C.9)]~\cite{IOH}. So we may take the
vacuum $\left( \begin{array}{c} 0 \\ |\la \ra \end{array} \right)$, with
$\b_0|\la \ra=0$, and create the representatives of nontrivial
cohomology classes by using the same operators as in the NS sector
given in theorem 4 for the cases (1) and (2) (but with half-odd-integers
$J$ and $m$). For case (3), we should take the vacuum $\left(
\begin{array}{c} |\la \ra \\ 0 \end{array} \right)$. Of course,
the mode expansions should be modified accordingly. It is interesting
to find that the modified oscillators in Ref.~\cite{IOH} (cf. (C.10))
appear automatically from this procedure.

For the study of the ground ring as well as correlation functions for
this system, it is desirable to find vertex operators to create the
states in the R sector on the conformal vacuum by using spin fields, since
we expect that the generating elements of the ground ring are in the R
sector. Here we present a calculation of OPE among the operators in
(4.10).\footnote{These results are obtained in Ref.~\cite{OST,BS}.} This
confirms our expectation that similar ground ring exists in the 2D
supergravity theory. We will perform the calculation in the different
($q=0$) picture in which the states in (4.10) are created by the
operators
\EQ
\Psi^{(+)}_{Jm} (z) = (J^-_0)^{J-m}
[J\psi^M(z)-i(1-J)\psi^L(z)] e^{iJ \phi^M(z)} e^{(1-J) \phi^L(z)}.
\EN

The operator algebra for (4.11) can be written as
\EQ
\Psi^{(+)}_{j_1,m_1}(z)\Psi^{(+)}_{j_2,m_2}(0)=\cdots +\frac{1}{z}
\sum_{j_3,m_3}F^{j_3,m_3}_{j_1,m_1,j_2,m_2}\Psi^{(+)}_{j_3,m_3}(0)+\cdots,
\EN
As in the bosonic case, the $SU(2)$ invariance requires the structure
constants $F$ to be of the form
\EQ
F^{j_3,m_3}_{j_1,m_1,j_2,m_2}=C^{j_1+j_2-1,m_1+m_2}_{j_1,m_1,j_2,m_2}
g(j_1,j_2),
\EN
where $C$ are the Clebsch-Gordan coefficients and $g(j_1,j_2)$ is a
function of $j_1$ and $j_2$ only. For $j_3=j_1+j_2-1$, $C$ are given by
\bea
C^{j_1+j_2-1,m_1+m_2}_{j_1,m_1,j_2,m_2}
&=&\frac{N(j_3,m_3)}{N(j_1,m_1)N(j_2,m_2)}\frac{j_2m_1-j_1m_2}
{\sqrt{j_3(j_3+1)}}, \nonumber\\
N(j,m)&=&\sqrt{\frac{(j+m)!(j-m)!}{(2j-1)!}}.
\ena

In order to determine $g(j_1,j_2)$, we consider the OPE for $m_1=j_1-1$
and $m_2=j_2$:
\EQ
\Psi^{(+)}_{j_1,j_1-1}(z)\Psi^{(+)}_{j_2,j_2}(0)=\cdots +\frac{1}{z}
F^{j_1+j_2-1,j_1+j_2-1}_{j_1,j_1-1,j_2,j_2}\Psi^{(+)}_{j_1+j_2-1,
j_1+j_2-1}(0)+\cdots .
\EN
We then compute the left hand side to find
\bea
&&\sqrt{2j_1}\Psi^{(+)}_{j_1,j_1-1}(z)\Psi^{(+)}_{j_2,j_2}(0) \nonumber\\
&=& \oint \frac{du}{2\pi i}\psi^M(u+z):e^{-i\phi^M(u+z)}:
:[j_1\psi^M(z)-i(1-j_1)\psi^L(z)]e^{ij_1\phi^M(z)+(1-j_1)\phi^L(z)}:
 \nonumber\\ && \times
:[j_2\psi^M(0)-i(1-j_2)\psi^L(0)]e^{ij_2\phi^M(0)+(1-j_2)\phi^L(0)}:
 \nonumber\\
&=& -\oint \frac{du}{2\pi i}\left[ j_1u^{-1-j_1}(u+z)^{-j_2}z^{j_1+j_2-1}
\{ j_2\psi^M(0)-i(1-j_2)\psi^L(0) \} \right. \nonumber\\
&& + j_2u^{-j_1}(u+z)^{-1-j_2}z^{j_1+j_2-1}
\{ j_1\psi^M(z)-i(1-j_1)\psi^L(z)\}  \nonumber\\
&& \left. +(j_1+j_2-1)u^{-j_1}(u+z)^{-j_2}z^{j_1+j_2-2}\psi^M(u+z)\right]
 e^{i(j_1+j_2-1)\phi^M-(j_1+j_2-2)\phi^L}.
\ena
Changing the integration variable as $u=xz$, the integration can be
performed to give
\EQ
\sqrt{2j_1}\Psi^{(+)}_{j_1,j_1-1}(z)\Psi^{(+)}_{j_2,j_2}(0)=
-\frac{1}{z}\frac{\Gamma(j_1+j_2)}{\Gamma(j_1)\Gamma(j_2)}
\Psi^{(+)}_{j_1+j_2-1,j_1+j_2-1}(0),
\EN
which is very similar to the bosonic case~\cite{KP}. Comparing this with
eqs.~(4.13) and (4.15), we find
\EQ
g(j_1,j_2)=\frac{\sqrt{j_1+j_2}(j_1+j_2-1)!}{\sqrt{2j_1j_2}(j_1-1)!
(j_2-1)!}.
\EN
After appropriate choice of normalization, this leads to the operator
algebra isomorphic to the $W_{\infty}$ or area-preserving diffeomorphism:
\EQ
\Psi^{(+)}_{j_1,m_1}(z)\Psi^{(+)}_{j_2,m_2}(0)=
\frac{1}{z}(j_2m_1-j_1m_2)\Psi^{(+)}_{j_1+j_2-1,m_1+m_2}(0).
\EN
which is identical to the bosonic 2D gravity theory. We can similarly
analyse the OPE for other operators in (4.7).

\sect{Discussions}

There are couple of important questions to be addressed.

Our understanding of the relation between the matrix models and
continuum approach has been improved recently, mainly because a method
to calculate some correlation functions in a continuum approach has been
found~\cite{GLI}.  However, there is still some gap in the
identification of the scaling operators in the matrix models and
the states in the
continuum approach. In the latter, we encounter ``non-decoupling of
null states"~\cite{JOE}. On the other hand, the presence of null states
{\it is} responsible for discrete states with higher ghost numbers
after applying the Felder's cohomology~\cite{FEL}. This aspect must
also be related to the question  what is the ground ring when $c^M<1$. A
thorough comparison of the results from two approaches will be important.

We would like to see the algebra associated with the discrete states
in the supersymmetric case. As explicitly shown in sect.~4, the discrete
states in theorem 4 satisfy the same area-preserving diffeomorphism
as the bosonic case~\cite{OST,BS}. This is an indication that the ground
ring is identical to the bosonic case. It is, however, necessary to
examine the structure in the R sector in order to identify the complete
structure of the ground ring in the supersymmetric case.

In a study of the ground ring, Witten has identified the matrix model
potential starting from a field theoretical approach~\cite{W}. It would
be interesting to see if we could find out a corresponding
supersymmetric matrix model in this manner.

It would be appropriate to mention some results of related papers. In
Ref.~\cite{BMP}, the authors studied the BRST cohomology for the minimal
models coupled to gravity. By imposing  Felder's cohomology~\cite{FEL},
they recovered the results by Lian and Zuckerman~\cite{LIZ}. In a recent
paper~\cite{BMP2}, the same authors have reported the cohomology
associated with the boundary states of conformal grids, which must be
relevant to the idetification of missing states in~\cite{LIZ} in
comparison with the matrix models. The states with $N_{FP}= \pm 1$ are
also discussed in a different approach~\cite{MUKHI}.

While preparing this report, we have received Ref.~\cite{BMP3} which
discusses the ground ring in the 2D supergravity theory.

\vspace{1cm}
\noindent{\em Acknowledgements}

One of us (K.I.) would like to thank A. Jevicki, B. Sazdovi\'c,
B. Uro{\u s}evi\'c for many discussions on related subjects.
He is also grateful to the Physics Department of Brown University
for its warm hospitality and financial support. The other (N. O.) would
like to acknowledge useful discussions with H. Suzuki and H. Tanaka.

\newpage
\appendix

{\Large\bf Appendix}

\sect{Hermiticity properties of Fock spaces}

We would like to discuss the hermiticities and inner products in the
charged Fock spaces assumed in our text implicitly. Some properties are
well known and may be found in some other literature~\cite{FEL,KMA}.
However they have not been discussed clearly and in a coherent manner
in the context of the present subject.

We will present our discussion on the bosonic case (see sect.~3 for the
notation), but the extension to supersymmetric case is straightforward.

For a generic value of $\la$, we find that the Virasoro generators
transform under the Fock-hermitian conjugation as
\EQ
L_n^{\dagger} (\la) = L_{-n} ( - \la^*),
\EN
which is the right hermiticity only for representations with $\la
\in i {\bf R}$ (pure imaginary). So for the gravity sector, the Fock
space hermiticity is the right one.  Obviously, we have to define
another hermiticity for the matter sector where $\la$ is real,
$c \le 1$. This is provided by using the parity operator which changes
signs of all the operators: $P (q, p, \phi_n) P = - (q, p, \phi_n)$;
its action on a Fock space is defined as $P {\cal F}(p) = {\cal F}(-p)$.
We then define $t$-operation by
\EQ
L_n^t(\la) = P  L_n(\la)^{\dagger}P = L_{-n}(\la).
\EN
This $t$-operation may be regarded as appropriate hermitian conjugation
for the case of $c \le 1$.\footnote{This possibility is pointed out
to one of the authors (K.I.) by M. Kato.}

The inner products which respect the hermiticity properties are given
by the following bilinear products defined over dual vector
spaces~\cite{FEL} for $c \le 1$,
\EQ
\lan y', y \ra \equiv |y'\ra^t |y\ra,
\EN
where $|y\ra \in {\cal F}(p)$ and $|y'\ra \in {\cal F}(-p)$;
for $c \ge 1$
\EQ
\lan y', y \ra \equiv |y'\ra^{\dagger} |y\ra
\EN
where $|y\ra \in {\cal F}(p)$ and $|y'\ra \in {\cal F}(p^*)$.
Note that the products are made compatible with $\lan \mu|\nu \ra
= \d_{\mu^*,\nu}$, which follows from $p^{\dagger}=p$. Here
$|\nu \ra$ is a state with a momentum eigenvalue $\nu$.

Some comments are in order.  As explained in the above, one may use
either Fock space hermiticity or the t-operation for $c=1$ case and both
of them provides us with the desired hermiticity of Virasoro generators.
If we use the t-operation as the hermitian conjugation, the $SU(2)$
currents no longer have definite hermiticity. Rather, we find
$sl(2,R)$ current algebra satisfied by modified (anti-hermitian)
currents, $j^0(z)= - J^0(z)$ and $j^{\pm}(z)=iJ^{\pm}(z)$.

\sect{$c^M<1$ CFTs coupled to gravity}

In the text, we have presented expressions for operators which create
states in $Ker Q_B /Im Q_B$ for $c^M=1$ (bosonic) or ${\hat c^M}=1$
(supersymmetric case). Here we will give a procedure to obtain the
states for $c^M<1$ bosonic theories starting from $c^M=1$.
One can easily extend the procedure to supersymmetric case.

The relations in the following discussion are most clearly seen if we
compactify the bosons for the matter as well as gravity sectors.
So let us first describe a compactified boson. For the boson on a torus
with a radius $R$, $\phi_L+\phi_R \equiv \phi_L+\phi_R + 2 \pi R {\bf Z}$,
the momentum and winding modes are quantized. It has been noticed by
many authors that the formula for chiral momenta may be compared with
(3.6): $p_L = t_{(j,k)}$ and $p_R = t_{(j,-k)}$. Here we understand
that $t_+= 2/R$ and $t_-=-R$. From these relations, we obtain $\la
(R)= \frac{1}{2}(R-2/R)$ and $c(R)= 1-6(R/{\sqrt 2}-{\sqrt 2}/R)^2$.
We thus see that $R={\sqrt 2}$ corresponds to $c=1$ theory.

As the gravity sector, we take a compactified boson whose
compactification radius is related to that of matter part by $R^L=iR^M$.
This relation is due to the anomaly cancellation (3.5). For later
convenience, let us parametrize the matter radius as $R^M={\sqrt 2}
e^{\omega}$, then $t^M_{\pm}(\omega)=\pm i t^L_{\pm}(\omega)= \pm
{\sqrt 2} e^{{\mp}\omega}$, and $c^M(\omega)=1-24 \sinh^2 \omega$.

Our strategy is to change the radius $R^M$ (or $\omega$) and relate
systems with different $c^M$. Rather than changing the parameter by hand,
we use a ``Lorentz" transformation defined by
\EQ
G(\omega){\hat o}G(-\omega)= \Omega(\omega){\hat o}
\EN
where
\EQ
\Omega(\omega)= \pmatrix{\cosh \omega & i \sinh \omega \cr
                         -i \sinh \omega & \cosh \omega \cr}
\EN
and ${\hat o}= q,~ p$ or $\phi_n$ are two-dimensional vectors; for
example, $q \equiv \left( \begin{array}{c} q^M \\ q^L  \end{array}
\right)$. The generator of the transformation is
\bea
G(\omega)&=& e^{\omega {\cal G}}, \nonumber\\
{\cal G} &=& q^M p^L- q^L p^M
-i\sum_{n\neq 0}\frac{1}{n}\a^M_{-n}\a^L_n.
\ena
The BRST charge of the system with $c^M(\omega)$ is obtained by
$Q_B(\omega)=G(\omega)Q_B(0) G(-\omega)$, which then implies the
isomorphism; $Ker Q_B(\omega)/Im Q_B(\omega)\sim Ker Q_B(0)/Im Q_B(0)$.

Let us examine this isomorphism in more detail. For
case (1) in theorem 1, we may conclude that our transformation provides
us with the isomorphism from the following observations: the relation
of the vacuum, $|\la(\omega)\ra\equiv|\la^M(\omega)\ra \otimes
|\la^L(\omega)\ra \otimes c_1 |0\ra_{gh}= G(\omega)|\la(0)\ra$;
$G(\omega)V(t_{(j,k)}^M(0),z) V(t_{(j,k)}^L(0),z)G(-\omega)=
V(t_{(j,k)}^M(\omega),z) V(t_{(j,k)}^L(\omega),z)$. The states for
$c^M=1$, which represent nontrivial cohomology classes in (2)
and (3) of the theorem 1, have nonvanishing bilinear products, discussed
in appendix A, in the following combinations:
the states with $N_{FP}=0$ in (2) and (3);
the states with $N_{FP}=1$ in (2) and those with $N_{FP}=-1$ in (3).
The states corresponding to (2) and (3) of theorem 1
for $c^M(\omega)$ may be obtained
from those in theorem 2 as follows. First use the operators listed in
theorem 2 and obtain states for $c^M=1$ system, and then act $G(\omega)$
on them. The resultant states are in $Ker Q_B(\omega)$ but not in
$Im Q_B(\omega)$, since the inner products are nonzero. [Actually the
inner products do not change under the transformation owing to
$P^M G(\omega)^{\dagger} P^M = G(-\omega)$.]

\sect{Proof of Theorems 1 and 3}

We are going to sketch the BRST analysis in our earlier paper~\cite{IOH},
treating NS, R sectors and bosonic theory in as parallel as possible.

As a preparation, let us define some notations:
\EQ
P^\pm (n) = \frac{1}{\sqrt 2}[(p^M +n \la^M)\pm i(p^L+n\la^L)]\equiv
\cases{\frac{1}{\sqrt 2}t^M_+(j-n) \cr \frac{1}{\sqrt 2}t^M_-(k-n) \cr};
\EN
and the light-cone variables
\bea
p^\pm &=& P^\pm(0)=\frac{1}{\sqrt 2}(p^M\pm i p^L),\;\;\;
q^\pm  = \frac{1}{\sqrt 2}(q^M\pm i q^L),\;\;\; \nonumber\\
\a^\pm_n &=&\frac{1}{\sqrt 2}(\a^M_n \pm i \a^L_n),\hspace{7mm}
\psi^\pm_r = \frac{1}{\sqrt {2}} (\psi^M_r\pm i\psi^L_r).
\ena
In eq.~(C.1), we have used the linearity of $P^\pm (n)$ in $n$, and
defined $j$ and $k$ as their zeros and
\EQ
t^M_{\pm}=\cases{
-\la^M\pm\sqrt{(\la^M)^2+2} & \mbox{for bosonic theory} \cr
-\la^M\pm\sqrt{(\la^M)^2+1} & \mbox{for supersymmetric theory} \cr }.
\EN
$j$ and $k$ are not necessarily integers at this stage. Note that (C.1)
is rewritten as (3.6) for $p^{M,L}$ and
\EQ
p^+p^-=P^+(0)P^-(0)=\shalf t^M_+t^M_-jk=\cases{-jk &
\mbox{for bosonic theory}
 \cr -\shalf jk & \mbox{for supersymmetric theory} \cr }.
\EN

The BRST charge has an expansion with respect to ghost zero modes as
\EQ
Q_B= c_0L-b_0M+d ~~~(~-\frac{1}{2}\c_0 F + 2\b_0K-\frac{1}{4}b_0\c^2_0~)
\EN
for NS sector and bosonic theory (for R sector, we add three other
terms in the bracket). Here
\EQ
L=p^+p^- +{\hat N}
\EN
with ${\hat N}$ as the level counting operator.

Our physical states are defined as nontrivial states satisfying
\EQ
Q_B|{\rm phys}\ra=0.
\EN
Using the relation $L=\{b_0,Q_B\}$ on a physical state, we have
\EQ
L|{\rm phys} \ra =Q_B b_0|{\rm phys}\ra,
\EN
which implies that the state is $Q_B$-trivial unless it is on-shell $L=0$.

The relative cohomology is defined as the cohomology of $Q_B$ on
${\cal F}(p^M, p^L) \cap Ker (b_0)$ $[\cap Ker (\b_0)]$ for the NS
sector and bosonic theory (R sector).
So the physical state condition becomes $L|{\rm phys}\ra=d|{\rm phys}\ra
=0$ ($F|{\rm phys}\ra=d|{\rm phys}\ra=0$) for the NS sector and
bosonic theory (R sector). Note
that since $F^2=L$, $V_F \equiv \{ |\psi\ra: F|\psi\ra=0 \} \subset
V_L \equiv \{ |\psi\ra: L|\psi\ra=0 \}$ in the R sector. This means
that any physical state is an on-shell state. In terms of the momenta
$p^+$ and $p^-$, the on-shell condition for both sectors (bosonic case)
is written as $p^+p^-+{\hat N}=-jk/2+{\hat N}(=-jk+{\hat N})=0$.
Hence $jk$ must be a positive integer in order to satisfy the on-shell
condition.

We may consider two possibilities depending on whether we have oscillator
excitations. If we have no oscillator excitation, then $j=0$ or $k=0$,
which corresponds to the case (1) of theorems 1 and 3.  In the R sector,
we find spinor structures of states as follows:
\EQ
\left( \begin{array}{c}
1 \\ 0  \end{array} \right)\cdot
|p^M, p^L> \hspace{5mm} \mbox{for}\hspace{5mm} p^+=0; \hspace{1cm}
\left( \begin{array}{c}
0 \\ 1  \end{array} \right)\cdot
|p^M, p^L> \hspace{5mm} \mbox{for}\hspace{5mm} p^-=0.
\EN
where we have used $\psi_0^\pm \equiv \frac{1}{\sqrt 2}(\psi_0^M \pm
i\psi_0^L) \equiv \shalf (\sigma_1 \pm i\sigma_2)$.
These vacuum states satisfy $d=0$ condition trivially.

Let us discuss the other case, $jk \ne 0$. Since $p^+ \ne 0$, we may
define the following operators in the R sector:
\bea
{\tilde \a}_n^\pm &\equiv& \a_n^\pm + n\theta\psi_n^\pm,
 \hspace{1cm}
{\tilde \psi}_n^\pm \equiv \psi_n^\pm - \theta\a_n^\pm,
 \nonumber\\
{\tilde c}_n &\equiv& c_n-\theta\c_n, \hspace{1cm}
{}~~{\tilde b}_n \equiv b_n+n\theta\b_n, \nonumber\\
{\tilde \c}_n &\equiv& \c_n+n\theta c_n, \hspace{1cm}
{\tilde \b}_n \equiv \b_n-\theta b_n,
\ena
where $\theta = \psi_0^+/p^+$, ($\theta^2=0$). In order to discuss the
two sectors at the same time, let us use the same notation ${\tilde
\a}_n^\pm,{\tilde \psi}_{n+\d}^\pm, \cdots$ for $\a_n^\pm,
\psi_{n+1/2}^\pm, \cdots$ in the NS sector, and for
${\tilde \a}_n^\pm, {\tilde \psi}_{n}^\pm, \cdots$ in the R sector.
For the bosonic theory, we use the same notation as NS sector
but, of course, without the fermions as well as super-ghosts.

We assign the degrees to the mode operators as follows:
\bea
&& deg\left({\tilde \a}^+_n,{\tilde \psi}^+_{n+\d},
{\tilde c}_n, {\tilde \c}_{n+\d}\right) = +1, \nonumber\\
&& deg\left({\tilde \a}^-_n, {\tilde \psi}^-_{n+\d},
{\tilde b}_n,{\tilde \b}_{n+\d} \right) = -1,
\ena
and 0 to ground states. All the states then carry
definite degrees. The operator $d$ is expanded according to the degree
as $d=d_0+d_1+d_2$.

Our strategy is first to study $d_0$-cohomology classes and then examine
if they can be extended to $d$-cohomology classes. Let us explain the
procedure in some details. The lowest degree term of $d$ is given by
(the second term is absent for bosonic theory)
\EQ
d_0 = \sum_{n\neq 0} P^+(n){\tilde c}_{-n}{\tilde \a}^-_n
-\frac{1}{2}\sum_{n+\d \neq 0}P^+(2n+2\d){\tilde \c}_{-(n+\d)}
{\tilde \psi}^-_{n+\d}.
\EN
We now define the operator
\EQ
K \equiv \sum_{n\neq 0}{}' \frac{1}{P^+(n)}{\tilde \a}^+_{-n}
{\tilde b}_n +\sum_{n+\d \neq 0}{}' \frac{2n+2\d}{P^+(2n+2\d)}
{\tilde \psi}^+_{-(n+\d)}{\tilde \b}_{n+\d}
\EN
where the primes imply that we have excluded the sum over $n=j$ or/and
$n+\d=j/2$, since $P^+(j)=0$.

The operator defined by $\hat N'=\{d_0,K\}$ is the number counting
operator for ${\tilde \a}^+_{-n}, {\tilde \a}^-_n, {\tilde b}_n,$
${\tilde c}_{-n} (n \neq j), {\tilde \psi}^+_{-(n+\d)}, {\tilde
\psi}^-_{n+\d}, {\tilde \b}_{n+\d}, {\tilde \c}_{-(n+\d)},
(2n+2\d \neq j)$ (The latter four oscillators are absent for the bosonic
theory). We may conclude from this relation that a $d_0$-closed
state with any of the above-listed oscillators excited is $d_0$-exact.
This can be seen similarly to eq.~(C.8). Therefore $d_0$-nontrivial
states must be created by the oscillators absent on the list, at the
level of $jk/2$ ($jk$ for bosonic) to satisfy the on-shell condition.
In particular, if $j =0$ or $j$ is not a nonzero integer, then we have no
oscillators available and only the ground state is allowed.  After
choosing momentum to satisfy the on-shell condition, we see that the
ground state is already a representative of nontrivial $d$-cohomology.
This is the case (1) of theorems 1 and 3.

{\samepage
In the following, we assume that $j \in {\bf Z}_+$ or ${\bf Z}_-$.
Since $jk$ is a positive integer, $k \in {\bf Z}_+$ or ${\bf Z}_-$
accordingly. For the bosonic theory,
we find the $d_0$-nontrivial states as
\EQ
\left({\tilde \a}^+_{-j}\right)^{k}\mid j, k \ra \;\; ,\;\;
{}~~{\tilde c}_{-j}\left({\tilde \a}^+_{-j}\right)^{k-1}\mid j, k \ra
\EN
for $j,k >0$ and}

\EQ
\left({\tilde \a}^-_j\right)^{-k}| j, k \ra\;\;,
{}~~{\tilde b}_j\left({\tilde \a}^-_j\right)^{-k-1}| j, k \ra
\EN
for $j,k <0.$  Here the vacuum is
$|j,k \ra=|p^M=t^M_{(j,k)} + \la^M
\ra \otimes| p^L=t^L_{(j,k)} + \la^L \ra \otimes c_1 |0\ra_{gh}$.

For the supersymmetric case, we find that there are four possible
cases listed on table 1 depending on the values of $(j,k)$.
$$
\begin{array}{|c|c|c|c} \hline
\begin{array}{c} ~~~~j \\  k~~~  \end{array}
& {\rm odd}+2\d & {\rm even}+2\d        \\ \hline
{\rm even} & {\rm eqs.(C.14,15)} & {\rm (all~parents~or~daughters)}\\ \hline
{\rm odd}  & \mbox{(no~on-shell~states)} & {\rm eqs.(C.16,17)} \\ \hline
\end{array}
$$
{\centerline {\bf Table 1}}

When $(j,k)=({\rm odd}+2\d$, even), we find the $d_0$-nontrivial states
as eqs.~(C.14,15) with $k$ in the exponents replaced by $k/2$.
The vaccum for the NS sector takes the same form as the bosnic case;
but for the R sector it has a spinor structure, $|j,k \ra=|p^M=t^M_
{(j,k)} + \la^M \ra \otimes| p^L=t^L_{(j,k)} + \la^L \ra \otimes c_1
|0\ra_{gh} \otimes \left( \begin{array}{c}0 \\ 1 \end{array} \right)$.

It is easy to see that no on-shell state can be constructed out of
available oscillators if $(j,k)=({\rm odd}+2\d$, odd).

The situation is quite different from bosonic case if $j={\rm even}+2\d$.
We may construct many states with different degrees and ghost numbers.
The generic patterns are given on tables 2 for $k=$odd and 3 for
$k=$even. On the tables, all the states $|*\ra$ represent nontrivial
$d_0$-cohomology classes.
$$
\begin{array}{|c|c|c|c|c|} \hline
\begin{array}{c} ~~~~~N_{FP} \\ {\rm deg.~~~}  \end{array}
& k-1   & k-2   &       &       \\ \hline
k       &       |{\rm daughter}>&       |{\rm daughter}>
&       &       \\ \hline
k-1     &       &
\begin{array}{c}
|{\rm parent}> \\
|{\rm daughter}> \end{array}    &
\begin{array}{c}
|{\rm parent}> \\
|{\rm daughter}>\end{array}     &       \\ \hline
 &       &       |{\rm nontrivial}\ra    &
\begin{array}{c}
|{\rm parent}> \\
|{\rm nontrivial}\ra \end{array} &
|{\rm parent}>  \\ \hline
\end{array}
$$
{\centerline {{\bf Table 2}: $k=$odd}}

$$
\begin{array}{|c|c|c|c|c|} \hline
\begin{array}{c} ~~~~~N_{FP}    \\      {\rm deg.~~~}   \end{array}
& k-1   & k-2   &       &       \\ \hline
k       & |{\rm daughter}>      & |{\rm daughter}>      & & \\ \hline
k-1     &       &
\begin{array}{c}
|{\rm parent}> \\
|{\rm daughter}> \end{array} &
\begin{array}{c}
|{\rm parent}> \\
|{\rm daughter}>\end{array}     &       \\ \hline
 &       &       &
|{\rm parent}>  &
|{\rm parent}>  \\ \hline
\end{array}
$$
{\centerline {{\bf Table 3}: $k=$even} }

Now that we have finished the analysis of $d_0$-cohomology, we are in a
position to extend the results to $d$-cohomology. In order to construct
a state representing nontrivial cohomology class of $d$, we may start
from a state nontrivial with respect to $d_0$ and add terms of higher
degrees. This is due to the following Lemma~\cite{BMP}:

\vspace{5mm}
{\large\bf Lemma 1}

The lowest degree term in a state nontrivial with respect to $d$ may
always be chosen to represent a nontrivial cohomology of
$d_0$. $\bullet$
\vspace{5mm}

Furthermore, we have~\cite{BMP}:

\vspace{5mm}
{\large\bf Lemma 2}

If, for each ghost number $N_{FP}$, the cohomology of $d_0$
is nontrivial for at most one fixed degree $k$ independent of $N_{FP}$,
then the cohomologies of $d_0$ and $d$ are isomorphic. $\bullet$
\vspace{5mm}

For (bosonic as well as supersymmetric) critical strings and bosonic
non-critical strings, the isomorphism has been proved.  These lemmas
are enough to derive the results in sect.~3.

In the super-Liouville theory, the assumption holds for $j={\rm odd}
+2\d$ and we may construct $d$-nontrivial states starting from (C.14,15).
However, when $j={\rm even}+2\d$, the assumption for the isomorphism
does not hold as can be seen from the tables 2 and 3. However, we have
in this case:

\vspace{5mm}
{\large\bf Lemma 3}

If the action of $d$ on a $d_0$-nontrivial state produces another
$d_0$-nontrivial state, those
two states cannot give rise to $d$-nontrivial states. $\bullet$
\vspace{5mm}

If we find states related as $d|\zeta\ra=|\xi\ra$, obviously both of
them do not contribute to the spectrum. The states related this way are
called ``parents" and ``daughters" respectively in the context of the
BRST formalism.  By studying relations of $d_0$-nontrivial states under
$d$, it is easy to show that they are classified as ``parents",
``daughters" or $d$-nontrivial states in the pattern indicated in the
tables 2 and 3; a ``daughter" is created from a ``parent" at the
lower-right position on the tables.  Only when $k=$odd, we may obtain
$d$-nontrivial states starting from the following states:
\bea
{\tilde \psi}^+_{-j/2}({\tilde \a}^+_{-j})^{(k-1)/2}|j, k \ra,
\nonumber\\{}
[({\tilde \a}^+_{-j})^{(k-1)/2}{\tilde \c}_{-j/2}-j(k-1){\tilde c}_{-j}
{\tilde \psi}^+_{-j/2}({\tilde \a}^+_{-j})^{(k-3)/2}] |j, k \ra
\ena
for $j,k >0$ and
\bea
{\tilde \psi}^-_{j/2}
({\tilde \a}^-_{j})^{-(k+1)/2}| j, k \ra, \nonumber\\
{} [({\tilde \a}^-_{j})^{-(k+1)/2}{\tilde \b}_{j/2}-\shalf {\tilde b}_{j}
{\tilde \psi}^-_{j/2}({\tilde \a}^-_{j})^{-(k+3)/2}]|j, k \ra
\ena
for $j,k <0.$

The states for NS sector obtained above are in the $q=-1$ picture, and
we have given in theorem 4 the explicit vertex representations of the
states for ${\hat c}=1$ theory in this picture. The operators (4.11), on
the other hand, are written in the $q=0$ picture.

This completes the proof of the results presented in sect.~4.


\newpage
\begin{thebibliography}{99}
\bibitem{BDG} E. Br\'{e}zin and V. Kazakov, \PL{B236} (1990) 144;\\
              M. R. Douglas and S.Shenker, \NP{B335} (1990) 635;\\
              D. J. Gross and A. A. Migdal, \PRL{64} (1990) 127.
\bibitem{PGB} G. Parisi, \PL{B238} (1990) 209;\\
              D. J. Gross and N. Miljkovi\'{c}, \PL{B238} (1990) 217;\\
              E. Br\'{e}zin, V. Kazakov and Al. B. Zamolodchikov,
              \NP{B338} (1990) 637;\\
              P. Ginsparg and J. Zinn-Justin, \PL{B240} (1990) 333.
\bibitem{JEV} S. R. Das and A. Jevicki, \MPL{A5} (1990) 1639;\\
              A. Jevicki, to appear in the Proceedings of 20th Int.
              Conf. on Differential Geometric Methods in Theoretical
              Physics (New York, Jun 3-7 1991), PRINT-91-0358 (BROWN)
              and references therein;\\
              K. Demeterfi, A. Jevicki and J. P. Rodrigues, \MPL{A6}
              (1990) 3199.
\bibitem{GKN} D. J. Gross and I. Klebanov, \NP{B344} (1990) 475;\\
              D. J. Gross, I. Klebanov and M. Newmann, \NP{B350} (1990)333.
\bibitem{POL} A. M. Polyakov, \MPL{A6} (1991) 635.
\bibitem{POL1} A. M. Polyakov, \PL{103B} (1981) 207.
\bibitem{DDK} J. Distler and H. Kawai, \NP{B321} (1989) 509;\\
              F. David, \MPL{A3} (1989) 1651.
\bibitem{LIZ} B. H. Lian and G. J. Zuckerman, \PL{B254} (1991) 417;
              \PL{B266} (1991) 21.
\bibitem{BMP} P. Bouwknegt, J. M. McCarthy and K. Pilch, CERN preprint,
              CERN-TH.6162/91 (1991).
\bibitem{W}   E. Witten, IAS preprint, IASSNS-HEP-91/51(1991).
\bibitem{KP}  I. Klebanov and A. M. Polyakov, \MPL{A6} (1991) 3273.
\bibitem{IOH} K. Itoh and N. Ohta, Fermilab preprint,
              FERMILAB-PUB-91/228-T (1991).
\bibitem{BMP2} P. Bouwknegt, J. M. McCarthy and K. Pilch, CERN preprint,
              CERN-TH.6279/91 (1991).
\bibitem{JRO} A. Jevicki and J. P. Rodrigues, Brown preprint,
	      BROWN-HET-813 (1991);\\
	      E. Marinari and G. Parisi, \PL{B240} (1990) 375;\\
	      S. Bellucci, T. R. Govindrajan, A. Kumar and R. N. Oerter
	      \PL{B249} (1990) 49;
	      L. Alvarez-Gaum\'e and J. L. Manes, CERN-TH-6067/91;\\
	      A. Dabholkar, Rutgers preprint RU-91-20 (1991).
\bibitem{KUO} T. Kugo and I. Ojima, \PTPS{66} (1979) 1.
\bibitem{FF}  B. L. Feigin and D. B. Fuchs, Seminar on Supermanifolds
              No. 5, ed. D. Leites.
\bibitem{KMA} M. Kato and S. Matsuda, Adv. Studies in Pure Math.
              {\bf 16} (1988) 205.
\bibitem{DHO} E. D'Hoker, \MPL{A6} (1991) 745.
\bibitem{FRI} D. Friedan, in {\em Recent Advances in Field Theories and
              Statistical Mechanics}, 1982 Les Houches summer school, ed.
              J. B. Zuber and R. Stora, North-Holland (1984).
\bibitem{KPZ} A. M. Polyakov, \MPL{A2} (1987) 893;\\
              V. G. Knizhnik, A. M. Polyakov and A. B. Zamolodchikov,
              \MPL{A3} (1988) 819.
\bibitem{LIO} T. L. Curtright and C. B. Thorn, \PRL{48} (1982) 1309;\\
	      E. Braaten, T. Curtright and C. Thorn, \PL{118B} (1982) 115;
	      \AP{147} (1983) 365;\\
	      E. Braaten, T. Curtright, G. Ghandour and C. Thorn,
	      \PRL{51} (1983) 19; \AP{153} (1984) 147;\\
	      N. Seiberg, \PTPS{102} (1991) 319;\\
	      J. Polchinski, in {\em Strings '90}, ed. R. Arnowitt et al.,
	      World Scientific (1991).
\bibitem{DHK} J. Distler, Z. Hlousek and H. Kawai, \IJMP{A5} (1990) 375.
\bibitem{KOG} M. Kato and K. Ogawa, \NP{B212} (1983) 443.
\bibitem{IKU} K. Itoh, \NP{342} (1990) 449;\\
              T. Kuramoto, \PL{B233} (1989) 363.
\bibitem{WIN} I. Bakas, \PL{B228} (1989) 57;\\
	      C. Pope, L. Romans and X. Sen, \NP{B339} (1990) 191;\\
	      E. Bergshoeff, M. P. Blencowe and K. S. Stelle, \CMP{128}
	      (1990) 213.
\bibitem{WY}  M. Wakimoto and H. Yamada, Hiroshima Math. J. {\bf 16}
              (1986) 427.
\bibitem{OHT} N. Ohta, \PR{D33} (1986) 1681; \PL{B179} (1986) 347.
\bibitem{MIT} M. Ito, T. Morozumi, S. Nojiri and S. Uehara, \PTP{75}
              (1986) 934.
\bibitem{ITO} K. Itoh, \IJMP{A6} (1991) 1233.
\bibitem{OST} N. Ohta, H. Suzuki and H. Tanaka, unpublished(1991).
\bibitem{BS}  B. Sazdovi\'{c}, private communication.
\bibitem{GLI} M. Goulian and M. Li, \PRL{66} (1991) 2051;\\
              P. Di Francesco and D. Kutasov, \PL{B261} (1991) 385;\\
              Y. Kitazawa, \PL{265B} (1991) 262;\\
              N. Sakai and Y. Tanii, \PTP{86} (1991) 547;\\
              Vl. S. Dotsenko, Paris VI preprint, PAR-LPTHE 91-18 (1991).
\bibitem{JOE} J. Polchinski, \NP{B357} (1991) 241.
\bibitem{FEL} G. Felder, \NP{B317} (1989) 215.
\bibitem{MUKHI} S. Mukhi, TIFR/TH/91-50 (1991).
\bibitem{BMP3} P. Bouwknegt, J. M. McCarthy and K. Pilch, CERN preprint,
              CERN-TH.6346/91 (1991).

\end{thebibliography}
\end{document}